\documentclass[journal,10pt]{IEEEtran} 
\usepackage[figurename=Fig.]{caption}
\usepackage{tikz}
\usetikzlibrary{calc, positioning, shapes.geometric, decorations.pathreplacing}
\usepackage{array}
\usepackage{verbatim}
\usetikzlibrary{calc, shapes, arrows, positioning}
\usepackage{authblk}
\usepackage{blindtext}
\usepackage{ifpdf}
\usepackage{graphicx}
\usepackage{tabulary}
\usepackage{amsmath,amsthm,amssymb}
\usepackage{amsmath}
\usepackage{kantlipsum}
\allowdisplaybreaks
\usepackage{cleveref}
\usepackage{lipsum}%
\usepackage{optidef}
\usepackage[english]{babel} 
\theoremstyle{plain}

\crefname{theorem}{Theorem}{theorem}
\crefname{lemma}{Lemma}{Lemmas}

\usepackage{cite}
\usepackage{dsfont}
\usepackage[justification=centering]{caption}
\usepackage{color}
\usepackage{algorithm}
\usepackage{algorithmicx, algpseudocode}
\usepackage{etoolbox}
\usepackage{subcaption}
\usepackage{balance}
\usepackage{empheq,etoolbox}
\usepackage[utf8]{inputenc}
\usepackage{multirow}
\usepackage{float}

\usepackage{amssymb}
\usepackage{pifont}
\usepackage[hmargin=1.27cm,top=1.3cm,bottom=1.3cm]{geometry}

\usepackage{bbm}

\usepackage{physics}

\usepackage{babel}
\floatstyle{plaintop}
\restylefloat{table}
\patchcmd{\subequations}
\usepackage{lipsum} 
\allowdisplaybreaks

\tikzset{brace/.style={decorate, decoration={brace}},
 brace mirrored/.style={decorate, decoration={brace,mirror}},
}

\newcounter{brace}
\newcounter{arrow}
\begin{document}
 \captionsetup[figure]{name={Fig.},labelsep=period}

\title{Sum Rate Maximization for NOMA-Assisted Uplink Pinching-Antenna Systems}

\bstctlcite{IEEEexample:BSTcontrol}
\author{Ming Zeng, Ji Wang, Xingwang Li, Gongpu Wang, Octavia A. Dobre and Zhiguo Ding

\thanks{This work was supported in part by NSERC under Grants RGPIN-2021-02636 and CRC-2022-00115, and in part by FRQNT under Grant 341270. (Corresponding author: Ming Zeng.)}

    \thanks{M. Zeng is with Laval University, Quebec City, Canada (email: ming.zeng@gel.ulaval.ca).}
    
    \thanks{J. Wang is with Central China Normal University, Wuhan, China (e-mail: jiwang@ccnu.edu.cn).}

\thanks{X. Li is with Henan Polytechnic University, Jiaozuo, China (email: lixingwang@hpu.edu.cn).}

\thanks{G. Wang is with Beijing Jiaotong University, Beijing, China (email: gpwang@bjtu.edu.cn).}

\thanks{O. A. Dobre is with Memorial University, St. John’s, Canada (e-mail: odobre@mun.ca).}

\thanks{Z. Ding is with The University of Manchester, Manchester, UK (e-mail:zhiguo.ding@ieee.org).}









    
    }
\maketitle

\begin{abstract}
In this paper, we investigate an uplink communication scenario in which multiple users communicate with an access point (AP) employing non-orthogonal multiple access (NOMA). A pinching antenna, which can be activated at an arbitrary point along a dielectric waveguide, is deployed at the AP to dynamically reconfigure user channels. The objective is to maximize the system sum rate by jointly optimizing the pinching-antenna's position and the users' transmit powers. 
{\color{black}Two scenarios are considered: one without quality-of-service (QoS) guarantees, and the other with QoS guarantees. In the former case, users transmit at full power, and the antenna position is determined using the particle swarm optimization (PSO) algorithm. In the latter, an alternating optimization approach is adopted, where {\color{black}a low-complexity} solution is derived for power allocation, and a modified PSO algorithm is applied to optimize the antenna position. 
Numerical results show that the proposed pinching-antenna-assisted system significantly improves the sum rate compared to the conventional fixed-antenna architecture. Furthermore, the NOMA-based approach consistently outperforms its TDMA-based counterpart. Finally, the proposed PSO-based method achieves near-optimal performance, particularly when the QoS constraints are moderate.}

\end{abstract}

\IEEEpeerreviewmaketitle

\section{Introduction}
Recently, flexible-antenna systems, such as fluid-antenna systems, have received significant attention due to their capability to dynamically reconfigure wireless channels \cite{Wong_TWC21}. By optimizing antenna positions, flexible-antenna systems have demonstrated superior performance compared to conventional fixed-location-antenna counterparts \cite{Saeid_GC24, ahmadzadeh2025}. However, the physical displacement of antennas in these systems is typically constrained to only a few wavelengths, which limits their ability to establish line-of-sight (LoS) links.
This limitation is particularly critical in high-frequency bands, such as the millimeter-wave and terahertz spectrum, where the absence of LoS links severely degrades system performance \cite{Atsushi_22, zeng2025_WCM, ding2024, Xiao_WCL25, yang2025, liu2025pinching, Xiao_COMML2025}. Addressing this challenge, the concept of pinching-antenna systems was introduced by NTT DOCOMO \cite{Atsushi_22}. In their experimental demonstration, a plastic pinch was applied to a dielectric waveguide to enable the radiation of radio waves. By strategically positioning the pinch, strong LoS links can be established for users previously constrained to non-line-of-sight (NLoS) conditions, thereby significantly enhancing overall system performance.

In multi-user pinching-antenna systems, the signal transmitted through a given dielectric waveguide is inherently a superposition of the signals from all served users, thereby motivating the adoption of non-orthogonal multiple access (NOMA) techniques \cite{ding2024, wang2024, fu2025}. Specifically, the authors in \cite{ding2024} investigated NOMA-assisted pinching-antenna systems with the objective of maximizing the sum rate. Analytical results in \cite{ding2024} demonstrated that NOMA-assisted architectures outperform their orthogonal multiple access counterparts in terms of sum rate performance.
However, achieving optimal system performance relies on the ability to activate pinching antennas at arbitrary positions along the waveguide \textemdash \ a requirement that poses practical implementation challenges. To address this, \cite{wang2024} proposed a low-complexity and hardware-friendly approach, where pinching-antennas were pre-installed at discrete, fixed locations prior to transmission. During operation, a subset of these antennas can be selectively activated to serve the users. 
In contrast to \cite{ding2024} and \cite{wang2024}, which primarily focused on the sum rate maximization, the study in \cite{fu2025} addressed power minimization for NOMA-assisted pinching-antenna systems subject to each user's minimum data rate requirement, and proposed an iterative power allocation algorithm. Numerical evaluations in \cite{fu2025} confirmed the efficiency and superior power-saving performance of pinching antennas over conventional ones.

Note that the aforementioned studies \cite{ding2024, wang2024, fu2025} primarily focused on downlink transmission in pinching-antenna systems. To the best of our knowledge, only a limited number of works \textemdash \ namely \cite{hou2025, Tegos_2025} \textemdash \ explored uplink transmission. However, none of these uplink-focused studies have investigated the integration of NOMA into pinching-antenna systems.
To address this gap, this paper considers a NOMA-assisted uplink scenario incorporating a single activated pinching antenna, with the objective of maximizing the system sum rate through a joint optimization of user transmit power and antenna positioning. 
{\color{black}Two scenarios are considered: without quality-of-service (QoS) guarantees and with QoS guarantees. In the first case, users transmit at maximum power, and the antenna position optimization problem is reformulated into a generalized bell-shaped membership function, which is inherently non-convex. To effectively solve this problem, we employ the particle swarm optimization (PSO) algorithm due to its ability to handle non-convex optimization tasks. In the second case, we apply an alternating optimization approach, where {\color{black}a low-complexity} solution is first derived for power allocation, and a modified PSO algorithm is then used to determine the antenna position. Simulation results show that the proposed pinching-antenna-assisted system delivers a significant improvement in sum rate when compared to the conventional fixed-antenna architecture. Furthermore, the NOMA-based approach consistently outperforms its TDMA counterpart. Finally, the PSO-based method is demonstrated to achieve near-optimal performance, particularly when the QoS requirements are not overly restrictive.}

\begin{figure}[ht!]
\centering
\includegraphics[width=0.8\linewidth]{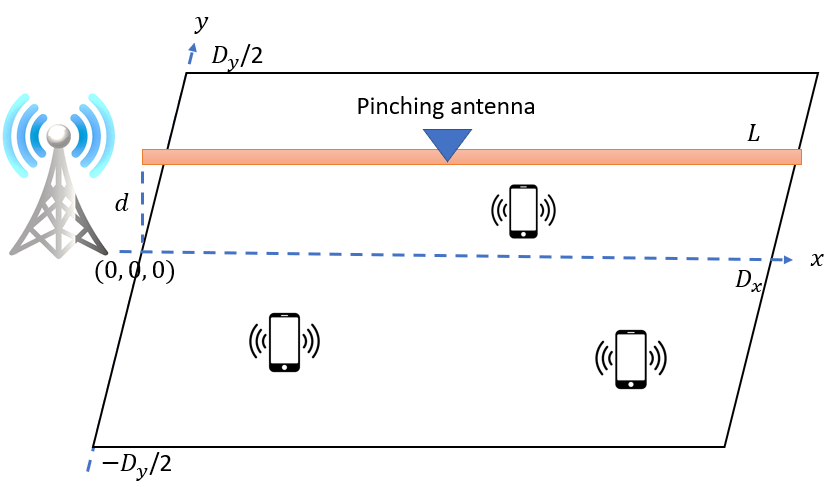}
\caption{{\color{black}Illustration of the considered uplink pinching-antenna system.}} 
\label{fig:Low_2}
\end{figure}

\section{System Model and Problem Formulation}
\subsection{System Model}
As illustrated in Fig. 1, we consider an uplink transmission scenario in which $M$ users communicate with a base station (BS) equipped with a single dielectric waveguide. We consider that a single pinching antenna is activated on the dielectric waveguide to establish LoS links with the users. Denote the user set by $\mathcal{M}=\{1, \cdot \cdot \cdot, M\}$. To model the spatial configuration, we adopt a three-dimensional Cartesian coordinate system. Without loss of generality, the waveguide is assumed to have a length of $L$ and is positioned parallel to the $x$-axis at a fixed height $d$, as shown in Fig. 1. The users are randomly distributed within a rectangular region lying in the $x-y$ plane, with dimensions $D_x$ and $D_y$. Let $\Phi^{\text{Pin}}=(x^{\text{Pin}}, 0, d)$ denote the location of the pinching antenna, while $\Phi_m= (x_m, y_m, 0)$ denote the coordinate of the $m$-th user, subject to $x_m \in [0, D_x]$ and $y_m \in [-D_y/2, D_y/2], \forall m \in \mathcal{M}$.


In this work, we consider a scenario in which all users are served simultaneously based on the principles of NOMA. At the BS, successive interference cancellation (SIC) is employed to mitigate inter-user interference. Without loss of generality, we assume that the users' signals are decoded in an ascending order. Then, the achievable data rate of the $m$th user is given by:

\begin{equation}
    R_m^{\text{NOMA}}=\log_2 \bigg(1+  \frac{\eta  P_m /\abs{\Phi_m - \Phi^{\text{Pin}}}^2 }{ \sum_{n=m+1} ^M \eta  P_n /\abs{\Phi_n - \Phi^{\text{Pin}}}^2  +  \sigma^2 } \bigg), 
\end{equation}
where $\eta= \frac{c^2}{16 \pi^2 f_c^2}$, with $c$ and $f_c$ denoting the speed of light and the carrier frequency, respectively. Additionally, $P_m$ represents the transmission power of the $m$th user, satisfying $P_m \leq P_m^{\max}$, with $P_m^{\max}$ denoting the maximum power constraint. $\sigma^2$ represents the power of additive white Gaussian noise at the BS.

\subsection{Problem Formulation}
In this paper, we aim to maximize the sum rate of the aforementioned uplink NOMA system by jointly optimizing the transmit power of the users and the position of the pinching antenna. The resulting optimization problem can be formulated as follows:
\begin{subequations} \label{P_NOMA}
   \begin{align}
    \max_{x^{\text{Pin}}, P_m}~ &\sum_{m=1}^M R_m^{\text{NOMA}} \\
    \text{s.t.}~& x^{\text{Pin}} \in [0, L], \\
    & P_m \leq P_m^{\max}, \forall m, 
\end{align} 
\end{subequations}
where (\ref{P_NOMA}b) limits the pinching antenna to the dielectric waveguide, while (\ref{P_NOMA}c) constrains the transmit power of each user to its maximum power.

\section{Proposed Solution}
Problem \eqref{P_NOMA} is non-convex due to the non-convex objective function (\ref{P_NOMA}a). To address it, we re-formulate (\ref{P_NOMA}a) as follows:
\begin{subequations} \label{NOMA_sum}
\begin{align}
    \sum_{m=1}^M R_m^{\text{NOMA}}      & =\sum_{m=1}^M  \log_2 \bigg(1+  \frac{\eta  P_m /\abs{\Phi_m - \Phi^{\text{Pin}}}^2 }{ \sum_{n=m+1} ^M \eta  P_n /\abs{\Phi_n - \Phi^{\text{Pin}}}^2  +  \sigma^2 } \bigg) \\
    &= \log_2 \bigg(1+  \frac{ \sum_{m=1} ^M \eta  P_m /\abs{\Phi_m - \Phi^{\text{Pin}}}^2   }{  \sigma^2 } \bigg),
\end{align}
\end{subequations}
where the final equality follows from the fact that the terms within the brackets in the sum-rate expression constitute a telescoping product \cite{Zeng_COMML21}. 

On this basis, problem \eqref{P_NOMA} can be re-written as
\begin{subequations} \label{P_NOMA_2}
   \begin{align}
    \max_{x^{\text{Pin}}, P_m}~ &\log_2 \bigg(1+  \frac{ \sum_{m=1} ^M \eta  P_m /\abs{\Phi_m - \Phi^{\text{Pin}}}^2   }{  \sigma^2 } \bigg) \\
    \text{s.t.}~& x^{\text{Pin}} \in [0, L], \\
    & P_m \leq P_m^{\max}, \forall m. 
\end{align} 
\end{subequations}

It can be readily verified that the objective function in (\ref{P_NOMA_2}a) is a monotonically increasing function of each user's transmit power. Therefore, to maximize the system sum rate, it is optimal for each user to transmit at its maximum power level, i.e., $ P_m =P_m^{\max}, \forall m \in \mathcal{M}$. With this rationale, the only remaining optimization variable is the position of the pinching-antenna, denoted by $x^{\text{Pin}}$. To facilitate further analysis, we remove the $\log(\cdot)$ function from the objective, leveraging its monotonicity, and reformulate the problem as:
\begin{subequations} \label{P_NOMA_3}
   \begin{align}
    \max_{x^{\text{Pin}} }~ &  \sum_{m=1} ^M   P_m^{\max} /\abs{\Phi_m - \Phi^{\text{Pin}}}^2     \\
    \text{s.t.}~& x^{\text{Pin}} \in [0, L].
\end{align} 
\end{subequations}

It is evident that problem \eqref{P_NOMA_3} shares the same optimal solution as the original formulation in \eqref{P_NOMA_2}, due to the monotonic nature of the logarithmic function. Furthermore, by substituting the expressions for $\Phi_m$ and $ \Phi^{\text{Pin}}$ with its corresponding coordinates, the objective function (\ref{P_NOMA_3}a) can be re-expressed as
\begin{equation}
     \sum_{m=1} ^M   \frac{P_m^{\max}}{\abs{x^{\text{Pin}} -x_m }^2   + y_m^2+ d^2},
\end{equation}
where $x_m$ and $y_m$ are the coordinates of the $m$th user in the $x$- and $y$-axis, which are assumed known.  

Each term in the objective function corresponds to a generalized bell-shaped membership function, which exhibits a symmetric bell-shaped profile \cite{Anuar_ISCAS19}. These functions are concave in the vicinity of their peak (i.e., near the center), convex in regions further from the center, and not globally convex \cite{Anuar_ISCAS19}.


\begin{figure}[ht!]
\centering
\includegraphics[width=0.8\linewidth]{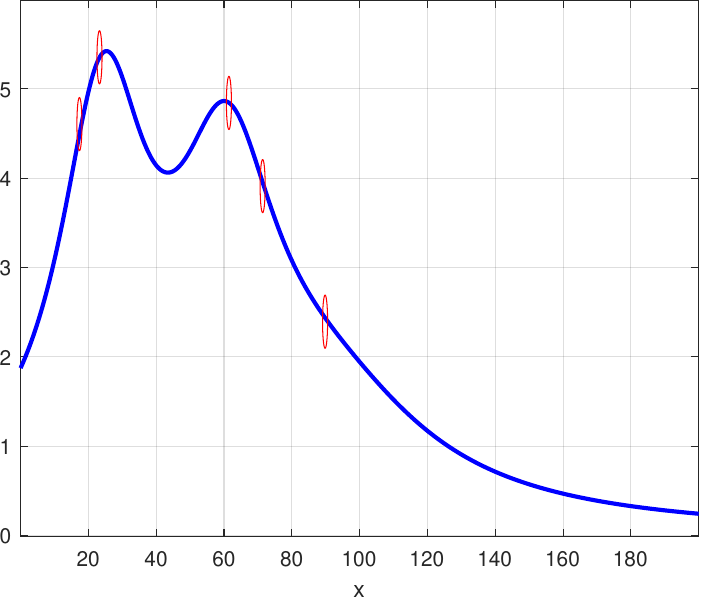}
\caption{One realization of the objective function with randomly generated values of $x_m$, $y_m$ and $d$, when $M=5$. The red circles denote the corresponding $x_m$ values.} 
\label{fig:Low_2}
\end{figure}

As shown in Fig. 2, the sum of multiple symmetric bell-shaped functions may exhibit multiple local maximums, making it challenging to derive a closed-form or analytical solution. Notably, multiple local maximums can occur even within the interval between two adjacent users along the $x$-axis. This characteristic renders traditional methods, such as the bisection algorithm, unsuitable for reliably identifying the maximum between two adjacent users.
Given that the problem involves a single optimization variable, a one-dimensional exhaustive search can always be employed to find the global optimum. However, the computational complexity of this approach increases with the search resolution (granularity).


As an alternative to the exhaustive search, we propose the use of PSO \textemdash \ a heuristic optimization technique that has demonstrated strong performance in identifying near-optimal solutions for non-convex problems \cite{Shi_Cat1988}.
The implementation of PSO in our context is as follows:
We first convert the original maximization problem into a minimization problem by negating the objective function. Let 
$I$ denote the total number of particles in the swarm. The position of the $i$th particle at iteration $t$ is represented by 
$x^{\text{Pin}}_i(t)$, and each particle is also associated with a velocity denoted by $v^{\text{Pin}}_i(t)$. The particle positions are iteratively updated, where the new position of each particle at the next iteration is computed as:
\begin{equation}
    x^{\text{Pin}}_i(t+1)=x^{\text{Pin}}_i(t)+ v^{\text{Pin}}_i(t+1).
\end{equation}
Simultaneously, the velocities of the particles are updated according to the following rule:
\begin{equation}
    v^{\text{Pin}}_i(t+1)=w v^{\text{Pin}}_i(t)+ c_1 r_1 ( pbest^i - x^{\text{Pin}}_i(t) ) + c_2 r_2 ( gbest - x^{\text{Pin}}_i(t) ),
\end{equation}
where $r_1$ and $r_2$ are random variables in the range $[0, 1]$; $w \in [0, 1]$ denotes the inertia weight, which controls the influence of a particle's previous velocity; and $c_1$ and $c_2$ are known as the cognitive and social acceleration coefficients, respectively. These parameters collectively govern the balance between exploration (global search) and exploitation (local refinement). The term $pbest^i$ represents the personal best position discovered by the $i$th particle, while $gbest$ is the global best position found by any particle in the swarm. Both are updated simultaneously at each iteration to reflect the best solutions encountered thus far. 
The algorithm terminates when either the maximum number of iterations is reached, or the relative change in the global best objective function values of two consecutive iterations falls below a predefined tolerance threshold.
In the context of our problem, the search space for $x^{\text{Pin}}$ is bounded. Specifically, the lower bound is set to $\min (x_m)$, and the upper bound is given by 
$\min (\max (x_m), L)$, since the values of $x^{\text{Pin}}$ outside this interval are guaranteed to yield suboptimal results due to their distances from all user positions.

{\color{black}
The computational complexity of the PSO algorithm is given by $O(I \times T \times C_f)$, where $T$ and $C_f$ denote the number of iterations and the computational cost of evaluating the objective function once, respectively. 
}

{\color{black}
\section{Guaranteed QoS for Each User}
The previous section addressed sum rate maximization in a pinching-antenna-aided multi-user scenario without enforcing individual rate constraints. In this unconstrained (greedy) setting, each user transmits at maximum power, which may lead to significant disparities in QoS among users. To address this limitation, we now consider a more fair formulation that maximizes the sum rate while ensuring a minimum rate requirement for each user. The resulting problem can be formulated as follows:
\begin{subequations} \label{P_NOMA_Rmin}
   \begin{align}
    \max_{x^{\text{Pin}}, P_m}~ &\sum_{m=1}^M R_m^{\text{NOMA}}\\
    \text{s.t.}~& x^{\text{Pin}} \in [0, L], \\
    & P_m \leq P_m^{\max}, \forall m, \\
    & R_m^{\text{NOMA}} \geq R_m^{\min}, \forall m,
\end{align} 
\end{subequations}
where (\ref{P_NOMA_Rmin}d) ensures the minimum rate for each user.


In this problem setting, transmitting at maximum power may no longer yield an optimal solution, as it can result in some users failing to meet their minimum rate requirements. Therefore, power control must be jointly considered alongside pinching antenna position optimization. However, the joint optimization of user transmit power and antenna position is inherently challenging due to the coupling between these variables. To tackle this complexity, we adopt an alternating optimization approach, where power allocation and antenna positioning are optimized iteratively, as detailed below.

\subsection{Power Allocation}
Under given pinching antenna position, the power allocation subproblem is given by
\begin{subequations} \label{P_NOMA_Rmin_PA}
   \begin{align}
    \max_{ P_m}~ &\sum_{m=1}^M \log_2 \bigg(1+  \frac{ \sum_{m=1} ^M \eta  P_m /\abs{\Phi_m - \Phi^{\text{Pin}}}^2   }{  \sigma^2 } \bigg) \\
    \text{s.t.}~& P_m \leq P_m^{\max}, \forall m, \\
    & \log_2 \bigg(1+  \frac{\eta  P_m /\abs{\Phi_m - \Phi^{\text{Pin}}}^2 }{ \sum_{n=m+1} ^M \eta  P_n /\abs{\Phi_n - \Phi^{\text{Pin}}}^2  +  \sigma^2 } \bigg) \geq R_m^{\min}, \forall m, 
\end{align} 
\end{subequations}
where it is assumed that the users channels are arranged in a descending order. Denote $h_m=\frac{ \eta }{ \abs{\Phi_m - \Phi^{\text{Pin}}}^2 \sigma^2} $ as the effective channel. Thus, we have $h_m>h_n, \forall m<n$. 

By removing the $\log(\cdot)$ function from the objective, as done previously, and reformulating the QoS constraints into linear expressions, we obtain the following problem:
\begin{subequations} \label{P_NOMA_Rmin_simplified}
   \begin{align}
    \max_{ P_m}~ &\sum_{m=1}^M  P_m h_m  \\
    \text{s.t.}~& P_m \leq P_m^{\max}, \forall m, \\
    & P_m h_m \geq (2^{R_m^{\min}}-1) \bigg( \sum_{n=m+1}^M P_n h_n+1 \bigg), \forall m. 
\end{align} 
\end{subequations}


The above problem is clearly a linear programming (LP) problem, and its optimal solution can therefore be efficiently obtained using standard LP solvers. To gain further analytical insight into the structure of the solution, we proceed to derive {\color{black}analytical} results as follows.

We begin by assigning the minimum required transmit power to each user such that its QoS constraint is just satisfied. This process can be performed sequentially, starting from the last user, i.e., user $M$, who experiences no interference. In this case, we have:
$P_M^{\min}=(2^{R_M^{\min}}-1)/h_M$. Then, we move to user $M-1$, whose power is given by $P_{M-1}^{\min}=(2^{R_{M-1}^{\min}}-1) \bigg( P_M^{\min} h_M+1 \bigg)/h_{M-1}= 2^{R_M^{\min}} \times (2^{R_{M-1}^{\min}}-1)/h_{M-1}$. Likewise, for any user $m$, we have 
\begin{equation}
    P_{m}^{\min}= 2^{ \sum_{n=m+1}^M R_{n}^{\min}} \times (2^{R_{m}^{\min}}-1)/h_{m}. 
\end{equation}

Note that if $P_{m}^{\min} > P_m^{\max}, \forall m$, 
the power allocation subproblem becomes infeasible. In the following analysis, we focus on the feasible case where all users can meet their minimum rate requirements within their respective power budgets.

After allocating the minimum required power to each user, there may be residual power available, which can be utilized to further improve the overall sum rate. This residual power can be allocated sequentially, starting from the first user (i.e., user 1). Examining the objective function reveals that the sum rate increases more significantly with the power allocated to users having stronger channel gains. Furthermore, due to SIC, users decoded earlier do not introduce interference to those decoded later. Therefore, it is optimal to allocate the remaining power in ascending order of user indices.

We begin with user 1. Since increasing $P_1$ directly improves the sum rate and does not affect the achievable rates of other users due to SIC, the optimal power allocation for user 1 is $P_1^{\text{Opt}}=P_1^{\max}$. Next, we consider user 2. To preserve the QoS of user 1, the power allocated to user 2 must satisfy the following condition {\color{black}according to (\ref{P_NOMA_Rmin_simplified}c)}:

\begin{subequations} \label{user 2}
    \begin{align}
      P_1^{\max} h_1 &\geq (2^{R_1^{\min}}-1) \bigg( P_2^{\text{Opt}} h_2+  \sum_{n=3}^M P_n^{\min} h_n+1 \bigg) \\
     \rightarrow P_2^{\text{Opt}} &\leq  \left( \frac{P_1^{\max} h_1 }{2^{R_1^{\min}}-1 }- \sum_{n=3}^M P_n^{\min} h_n-1 \right)/h_2. 
    \end{align}
\end{subequations}

If $ \left( \frac{P_1^{\max} h_1 }{2^{R_1^{\min}}-1 }- \sum_{n=3}^M P_n^{\min} h_n-1 \right)/h_2 <P_2^{\max}$, we have $P_2^{\text{Opt}} =  \left( \frac{P_1^{\max} h_1 }{2^{R_1^{\min}}-1 }- \sum_{n=3}^M P_n^{\min} h_n-1 \right)/h_2 $ {\color{black}to preserve the QoS of user 1. Moreover, the whole power allocation procedure terminates, as further power allocation to other users will lead to QoS violation of user 1.}
Otherwise, if the above condition is not satisfied, we set $ P_2^{\text{Opt}}= P_2^{\max}$ and proceed similarly to user 3. This iterative process continues either until an early termination condition is met or until all users have been considered. 

\subsection{Pinching Antenna Position Optimization}
Under given power allocation, the pinching antenna position optimization subproblem can be simplified as
\begin{subequations} \label{P_NOMA_3_Rmin}
   \begin{align}
    \max_{x^{\text{Pin}} }~ &  \sum_{m=1} ^M   P_m /\abs{\Phi_m - \Phi^{\text{Pin}}}^2     \\
    \text{s.t.}~& x^{\text{Pin}} \in [0, L], \\
    &  R_m^{\text{NOMA}} \geq R_m^{\min}, \forall m.
\end{align} 
\end{subequations}

The only difference between the above problem and problem \eqref{P_NOMA_3} lies in constraint (\ref{P_NOMA_3_Rmin}c). Therefore, the proposed PSO-based algorithm remains applicable for solving problem \eqref{P_NOMA_3_Rmin}, with appropriate modifications. To enforce the QoS requirements, we incorporate a penalty term into the objective function to penalize any violation of the minimum rate constraints. The modified objective function is given by:
\begin{equation}
    \sum_{m=1} ^M   P_m /\abs{\Phi_m - \Phi^{\text{Pin}}}^2 - \lambda \sum_{m=1}^M \mathbbm{1} ( R_m^{\text{NOMA}} - R_m^{\min})  , 
\end{equation}
where $ \lambda$ denotes the penalty factor, which is a large positive constant used to heavily penalize any violation of the QoS constraints. 
The function $\mathbbm{1}(\cdot)$ is the indicator function, which equals 1 if the condition inside the parentheses is true, and 0 otherwise. Based on this penalized objective, the previously proposed PSO algorithm can still be applied. 

\subsection{Alternating Optimization until Convergence}
The two steps described above---power allocation and pinching antenna position optimization---are performed iteratively in an alternating manner until convergence is achieved. To ensure convergence, the following stopping criterion is adopted: within the PSO algorithm, if the newly obtained heuristic solution does not improve upon the previous one, no update is made, and the entire alternating optimization process is terminated.
}

\begin{figure}[ht!]
\centering
\includegraphics[width=0.9\linewidth]{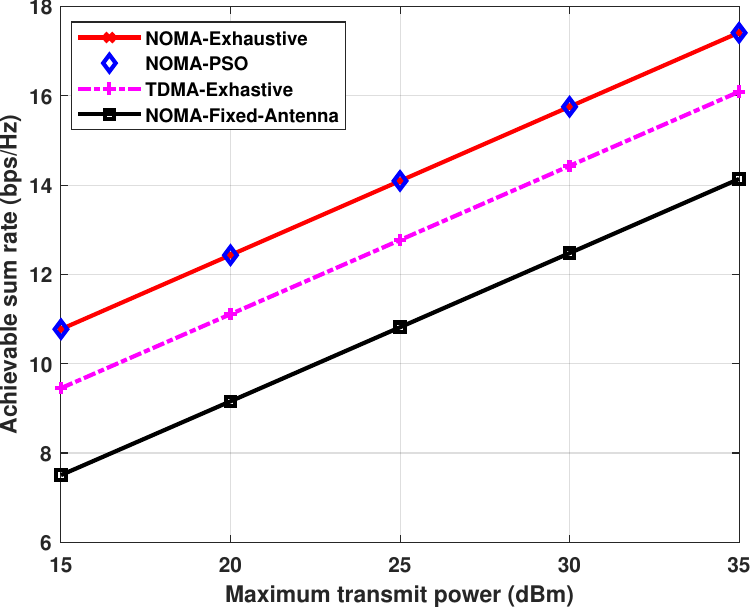}
\caption{Achievable sum rate versus the maximum transmit power constraint at the users when $R^{\min}=0.5$ bps/Hz.} 
\label{fig:power}
\end{figure} 

\section{Numerical Results}
{\color{black}Numerical simulations are carried out to evaluate the performance of the proposed pinching-antenna system (denoted as \textit{NOMA-PSO}) against several benchmarks. First, it is compared with a conventional fixed-antenna system (denoted as \textit{NOMA-Fixed-Antenna}), in which the antenna is placed at a fixed position of $(0, 0, d)$ meters. Additionally, two benchmark schemes are considered to assess performance bounds: (i) \textit{NOMA-Exhaustive}, representing the upper bound obtained by exhaustively searching all possible positions of the pinching antenna, and (ii) \textit{TDMA-Exhaustive}, which denotes the optimal TDMA solution for pinching-antenna system, also derived via exhaustive search.}
The default simulation parameters are as follows \cite{xie2025}: carrier frequency $f_c=28$ GHz, antenna height $d=3$ m, and noise power $\sigma^2=-90$ dBm. The number of users is set to 3, randomly deployed within a service area of $D_x=60$ m and $D_y=20$ m. 
Additionally, the length of the dielectric waveguide is set to $L=D_x$. All presented results are averaged over $10^2$ independent random realizations.


Figure 3 shows the achievable sum rate as a function of the users' maximum transmit power when $R^{\min}=0.5$ bps/Hz. As expected, the sum rate increases for all schemes with rising transmit power. 
The results clearly demonstrate that the pinching-antenna system significantly outperforms the fixed-antenna counterpart, benefiting from its capability to dynamically reposition the antenna closer to the users. Among the pinching-antenna configurations, the NOMA-based scheme achieves a higher sum rate than the TDMA-based one. 
Furthermore, the sum rate achieved by the proposed PSO-based optimization closely matches that obtained via an exhaustive search, confirming its near-optimal performance.

{\color{black}
Figure 4 illustrates the variation of the achievable sum rate as a function of the minimum rate requirement $R^{\min}$ at the users. Not surprisingly, the sum rate decreases for all considered schemes as $R^{\min}$ increases. A significant performance gap is observed between the pinching-antenna-assisted schemes and the conventional fixed-antenna scheme, further highlighting the advantages of the pinching antenna architecture. Among the evaluated methods, the NOMA-based schemes consistently outperform their TDMA counterpart, although this performance gap gradually narrows with increasing 
$R^{\min}$. Notably, when $R^{\min} \leq 1.1$ bps/Hz, the performance of \textit{NOMA-PSO} closely aligns with that of \textit{NOMA-Exhaustive}. However, as $R^{\min}$ increases beyond this threshold, a slight performance gap emerges, yet NOMA-PSO continues to demonstrate strong performance, maintaining a favorable trade-off between complexity and efficiency even under more stringent QoS requirements.
}

\begin{figure}[ht!]
\centering
\includegraphics[width=0.9\linewidth]{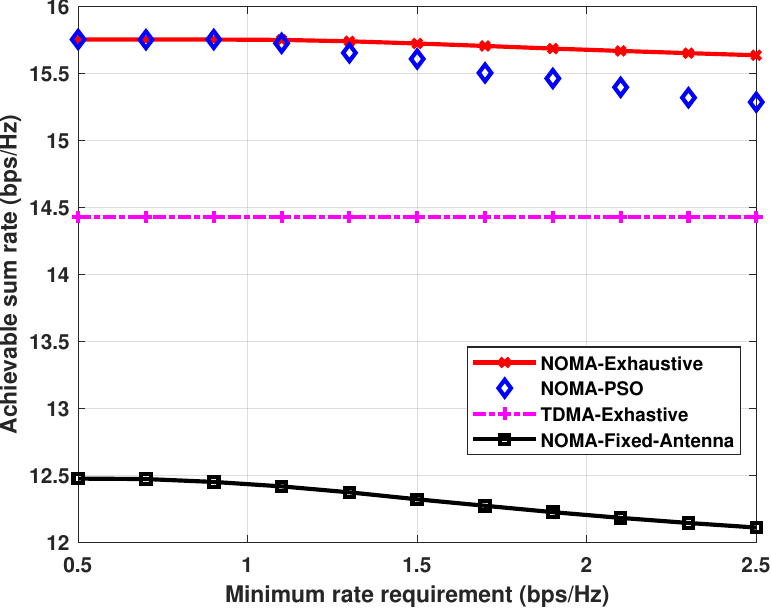}
\caption{Achievable sum rate versus $R^{\min}$ with $P^{\max}=30$ dBm.} 
\label{fig:distance}
\end{figure}

\section{Conclusion} 
\label{Sec:Conclusion}
{\color{black}In this paper, we studied sum rate maximization in an uplink NOMA system with a pinching antenna. For the case without QoS constraints, we showed the antenna optimization problem is non-convex and developed a PSO-based method to obtain near-optimal solutions. With QoS guarantees, we proposed an alternating optimization framework combining low-complexity power allocation and a modified PSO algorithm for antenna positioning. Numerical results demonstrated that pinching antennas significantly outperform fixed architectures, and that NOMA consistently outperforms TDMA. These results highlight the promise of pinching-antenna systems for enhancing spectral efficiency, with future work directed toward extending the analysis to systems with multiple pinching antennas and/or multiple waveguides.}

\bibliographystyle{IEEEtran}
\bibliography{biblio}

\begin{thebibliography}{10}
\providecommand{\url}[1]{#1}
\csname url@samestyle\endcsname
\providecommand{\newblock}{\relax}
\providecommand{\bibinfo}[2]{#2}
\providecommand{\BIBentrySTDinterwordspacing}{\spaceskip=0pt\relax}
\providecommand{\BIBentryALTinterwordstretchfactor}{4}
\providecommand{\BIBentryALTinterwordspacing}{\spaceskip=\fontdimen2\font plus
\BIBentryALTinterwordstretchfactor\fontdimen3\font minus \fontdimen4\font\relax}
\providecommand{\BIBforeignlanguage}[2]{{%
\expandafter\ifx\csname l@#1\endcsname\relax
\typeout{** WARNING: IEEEtran.bst: No hyphenation pattern has been}%
\typeout{** loaded for the language `#1'. Using the pattern for}%
\typeout{** the default language instead.}%
\else
\language=\csname l@#1\endcsname
\fi
#2}}
\providecommand{\BIBdecl}{\relax}
\BIBdecl

\bibitem{Wong_TWC21}
K.-K. Wong \emph{et~al.}, ``Fluid antenna systems,'' \emph{IEEE Trans. Wireless Commun.}, vol.~20, no.~3, pp. 1950--1962, Mar. 2021.

\bibitem{Saeid_GC24}
S.~Pakravan \emph{et~al.}, ``Robust resource allocation for over-the-air computation networks with fluid antenna array,'' in \emph{Proc. IEEE GC Workshops}, 2024, pp. 1--6.

\bibitem{ahmadzadeh2025}
\BIBentryALTinterwordspacing
M.~Ahmadzadeh \emph{et~al.}, ``Enhanced over-the-air federated learning using {AI}-based fluid antenna system,'' 2025. [Online]. Available: \url{https://arxiv.org/abs/2407.03481}
\BIBentrySTDinterwordspacing

\bibitem{Atsushi_22}
A.~Fukuda \emph{et~al.}, ``Pinching antenna using a dielectric waveguide as an antenna,'' \emph{Technical Journal}, vol.~23, no.~3, pp. 5--12, Jan. 2022.

\bibitem{zeng2025_WCM}
\BIBentryALTinterwordspacing
M.~Zeng \emph{et~al.}, ``Resource allocation for pinching-antenna systems: State-of-the-art, key techniques and open issues,'' 2025. [Online]. Available: \url{https://arxiv.org/abs/2506.06156}
\BIBentrySTDinterwordspacing

\bibitem{ding2024}
Z.~Ding, R.~Schober, and H.~V. Poor, ``Flexible-antenna systems: A pinching-antenna perspective,'' \emph{IEEE Trans. Commun.}, to appear in 2025.

\bibitem{Xiao_WCL25}
J.~Xiao \emph{et~al.}, ``Frequency-selective modeling and analysis for {OFDM}-integrated wideband pinching-antenna systems,'' \emph{IEEE Wireless Commun. Lett.}, pp. 1--1, 2025.

\bibitem{yang2025}
\BIBentryALTinterwordspacing
Z.~Yang \emph{et~al.}, ``Pinching antennas: Principles, applications and challenges,'' 2025. [Online]. Available: \url{https://arxiv.org/abs/2501.10753}
\BIBentrySTDinterwordspacing

\bibitem{liu2025pinching}
\BIBentryALTinterwordspacing
Y.~Liu \emph{et~al.}, ``Pinching-antenna systems (pass): Architecture designs, opportunities, and outlook,'' 2025. [Online]. Available: \url{https://arxiv.org/abs/2501.18409}
\BIBentrySTDinterwordspacing

\bibitem{Xiao_COMML2025}
J.~Xiao, J.~Wang, and Y.~Liu, ``Channel estimation for pinching-antenna systems ({PASS}),'' \emph{IEEE Commun. Lett.}, vol.~29, no.~8, pp. 1789--1793, 2025.

\bibitem{wang2024}
K.~Wang \emph{et~al.}, ``Antenna activation for {NOMA} assisted pinching-antenna systems,'' \emph{IEEE Wireless Commun. Lett.}, vol.~14, no.~5, pp. 1526--1530, 2025.

\bibitem{fu2025}
\BIBentryALTinterwordspacing
Y.~Fu, F.~He, Z.~Shi, and H.~Zhang, ``Power minimization for {NOMA}-assisted pinching antenna systems with multiple waveguides,'' 2025. [Online]. Available: \url{https://arxiv.org/abs/2503.20336}
\BIBentrySTDinterwordspacing

\bibitem{hou2025}
\BIBentryALTinterwordspacing
T.~Hou, Y.~Liu, and A.~Nallanathan, ``On the performance of uplink pinching antenna systems ({PASS}),'' 2025. [Online]. Available: \url{https://arxiv.org/abs/2502.12365}
\BIBentrySTDinterwordspacing

\bibitem{Tegos_2025}
S.~A. Tegos \emph{et~al.}, ``Minimum data rate maximization for uplink pinching-antenna systems,'' \emph{IEEE Wireless Commun. Lett.}, vol.~14, no.~5, pp. 1516--1520, 2025.

\bibitem{Zeng_COMML21}
M.~Zeng \emph{et~al.}, ``Sum rate maximization for {IRS}-assisted uplink {NOMA},'' \emph{IEEE Commun. Lett.}, vol.~25, no.~1, pp. 234--238, Jan. 2021.

\bibitem{Anuar_ISCAS19}
A.~Dorzhigulov and A.~P. James, ``Generalized bell-shaped membership function generation circuit for memristive neural networks,'' in \emph{Proc. IEEE ISCAS}, 2019, pp. 1--5.

\bibitem{Shi_Cat1988}
Y.~Shi and R.~Eberhart, ``A modified particle swarm optimizer,'' in \emph{Proc. Proc. IEEE Int. Conf. Evol. Comput., IEEE World Congr. Comput. Intell.}, 1998, pp. 69--73.

\bibitem{xie2025}
X.~Xie \emph{et~al.}, ``A low-complexity placement design of pinching-antenna systems,'' \emph{IEEE Commun. Lett.}, vol.~29, no.~8, pp. 1784--1788, 2025.

\end{thebibliography}

\balance

\end{document}